\documentclass{aa}
\usepackage{epsf}
\usepackage{amssymb}
\usepackage{graphics}
\usepackage[dvips]{epsfig}
\newcommand{\teff}{T_{\mathrm{eff}}}         
\begin{document}

\title{Binaries discovered by the SPY project. III. HE\,2209$-$1444: a
massive, short period double degenerate%
\thanks{Based on observations collected at the German-Spanish Astronomical
Center (DSAZ), Calar Alto, operated by the Max-Planck-Institute f\"ur
Astronomie jointly with the Spanish National Commission for Astronomie}
\thanks{Based on observations at the Paranal Observatory of the European
Southern Observatory for program No.\ 165.H-0588(A) and 167.D-0407(A) }
\thanks{Based on observations made with the INT operated on the island of 
La Palma by the Isaac Newton Group in the Spanish Observatorio del Roque de 
los Muchachos of the Instituto de Astrofisica de Canarias}
}
\author{C.A.~Karl\inst{1}
\and R.~Napiwotzki\inst{1}
\and G. Nelemans\inst{2}
\and N.~Christlieb\inst{3}
\and D. Koester\inst{4}
\and U.~Heber\inst{1}
\and D.~Reimers\inst{3}
}

\institute{Dr.-Remeis-Sternwarte, Astronomisches Institut der Universit\"at
  Erlangen-N\"urnberg, Sternwartstr.~7, 96049~Bamberg, Germany
\and Institute of Astronomy, Madingley Road, CB3 0HA, Cambridge, UK
\and Hamburger Sternwarte, Universit\"at Hamburg, Gojenbergsweg 112, 21029
     Hamburg, Germany
\and Institut f\"ur Theoretische Physik und Astrophysik, Universit\"at Kiel,
  24098 Kiel, Germany
}
\offprints{C.A. Karl (karl@sternwarte.uni-erlangen.de)}
\date{Received: date; accepted: date}
\titlerunning{Binaries discovered by the SPY project. III. HE\,2209$-$1444}
\authorrunning{Karl et al.}

\abstract{
In the course of our search for double degenerate (DD) binaries as potential
progenitors of type Ia supernovae with the UVES spectrograph at the ESO VLT
(ESO {\bf S}N Ia {\bf P}rogenitor surve{\bf Y} - SPY) we discovered
HE\,2209$-$1444 to be a double-lined system consisting of
two DA white dwarfs. From the analysis of the radial velocity curve 
we determined the period of the system
to be $P = 6^{\rm h}38^{\rm m}47^{\rm s}$. The semi-amplitudes 
for both indi\-vi\-dual components are 
109\,kms$^{-1}$ each.
A model atmosphere analysis enabled us to derive individual temperatures for 
both components (8490\,K and 7140\,K, resp.) and masses of $0.58M_\odot$ for
each component. The total mass of the system is $1.15\pm 0.07 M_\odot$.
The system will loose angular momentum due to gravitational wave radiation
and therefore will merge within 5\,Gyrs -- less than a Hubble time. 
HE\,2209$-$1444 is the second massive, short period double degenerate detected by
SPY.
Its total mass is about 20\% below the Chandrasekhar mass limit and
therefore it does not qualify as a potentional SN Ia progenitor. However,
together with our previous detections it supports the view that
Chandrasekhar mass systems do exist.
}
\maketitle

\section{Introduction}
Supernovae of type Ia (SN Ia) play an outstanding role in the study of cosmic evolution.
In particular, they are regarded as one of the best standard candles for the determination
of the cosmological parameters $H_0$, $\Omega$, and $\Lambda$ (e.g. Riess et
al.\ \cite{R98},
Leibundgut\ \cite{L01}). However, the nature of their progenitors remains a mystery
(e.g. Livio\ \cite{L00}).

There is general consensus that the SN event is due to the thermonuclear
explosion of a white dwarf likely when the Chandrasekhar limit of
$\approx$\,1.4\,$M_{\odot}$ is reached. But the exact nature of the progenitor
system is still unclear.  While it must be a binary with matter being
transferred to the white dwarf from a companion, two main scenarios exists. 
According
to the so-called double degenerate (DD) scenario (Iben and Tutukov\
\cite{I84}), the mass donating companion is also a white dwarf.
In the so-called singe
degenerate (SD) scenario (Whelan and Iben\ \cite{W73}), the mass do\-na\-ting
component is a red giant/subgiant.

A DD has to fulfill two criteria to be considered as a potential SN Ia
progenitor.  The total mass of the system has to exceed the Chandrasekhar
limit, and the two components have to merge in less than a Hubble time due to
the loss of angular momentum via gravitational wave radiation.  During the
last decades, several systematic radial velocity (RV) searches have been
undertaken (see Marsh\ \cite{M00} for a review). By now, combining all surveys
nearly 180 white dwarfs
have been checked for RV variations yielding a sample of 18 DDs
with periods of less than 6.3 days (Marsh\ \cite{M00}, Maxted et al.\
\cite{Metal00a}). Among these 18 systems, only 6 double-lined systems have
been found (in three of them, the companion is barely detectable). However,
none of the known 18 systems seems to be massive enough to qualify as a SN Ia
precursor. Since theoretical simulations suggest only a few percent of all
close DDs to be potential SN Ia progenitors (Iben et al.\ \cite{I97}, Nelemans
et al.\ \cite{NYP01}) this result is not surprising and a larger sample size
is needed to find a potential SN Ia progenitor.


Therefore\,--\,to perform a definite test of the DD scenario\,--\,we have
embarked on a large spectroscopic survey of $\approx$ 1000 white dwarfs
using the UV\,-\,Visual\,Echelle\,Spectrograph (UVES) 
at the ESO VLT UT2 (Kueyen), searching for RV variations of
white dwarfs
and pre white dwarfs (ESO {\bf S}N Ia {\bf P}rogenitor surve{\bf Y} - {\bf SPY}).
The ongoing SPY project already provided a wealth of new RV variable DDs (Napiwotzki
et al.\ \cite{N01a} and\ \cite{N03}).
In particular, more than a dozen double-lined systems have been found by
SPY. These are of outstanding importance because the radial velocity
curves can be solved for all system parameters if the gravitational
redshifts of the components are unequal (see Sect.~\ref{radvel_section}).
An analysis of the subdwarf B plus white dwarf
system HE\,1047$-$0436 was presented in the first paper of this series
(Napiwotzki et al.\ \cite{N01b}).  In paper II (Napiwotzki et al.\
\cite{N02}) we have presented the double-lined
system HE\,1414$-$0848, which has a system mass of 1.26
$M_{\odot}$, but will merge only in two Hubble times.

In this paper we report on the results of our follow-up spectroscopy of
the new double-lined system HE\,2209$-$1444. 
We will show that this system is another massive DD, but
with an orbital
period shorter than the HE\,1414$-$0848 system.
After a description of the observations and data reduction in
Sect.~\ref{obs_section} we deal with the
determination of the system's radial velocity curves and orbital
parameters in Sect.~\ref{radvel_section}. Section~\ref{s:analysis}
gives a model atmosphere analysis of the spectra.
We compare the results to predictions of population
synthesis models in Sect.~\ref{discussion} and
summarize in a final Sect.~\ref{summary}.

\section{Observations and data analysis} \label{obs_section}

HE\,2209$-$1444 ($\alpha_{\rm 2000}$\,=\,$22^{\rm h}12^{\rm m}18.1^{\rm s}$,
$\delta_{\rm 2000}$\,=\,$-14^{\rm o}29'48''$,
$B_{\mathrm{pg}}$\,=\,$15.3^{\rm mag}$)
was discovered by the Hamburg ESO survey (HES; Wisotzki et al.\ \cite{W00},
Christlieb et al.\ \cite{C01}) as a potential cool white dwarf and, therefore,
was included in our SPY project. From the first survey spectrum taken in
December 2000 we found HE\,2209$-$1444 to be a double-lined 
binary system consisting of two DA white dwarfs.
The H$_\alpha$ line cores of both components were separated by 4.5\,\AA {}
in the discovery spectrum, corresponding to a RV difference of 200\,km/s.

Therefore we observed the system during our follow-up observations
for the SPY project at the Calar Alto Observatory, Spain.
From 2001 July\,6$^{\mathrm{th}}$ to July\,15$^{\mathrm{th}}$ 
a total of 15 medium resolution 
spectra were obtained with
the TWIN spectrograph at the 3.5\,m telescope.
For the blue arm we used grating No.\,5 while No.\,6 was used in the
red. Both gratings have 1200\,lines/mm, giving a dispersion
of 0.54\,\AA/pix.
The wavelength coverage ranges from
3900\,\AA {} to 5000\,\AA {} at a resolution of 1.3\,\AA {} in
the blue part of the spectrum.
In the red part, the wavelength scale ranges from
6000\,\AA {} to 7000\,\AA {} at a resolution of 1.2\,\AA {}.
The data were reduced with a set of standard routines for long slit spectra
from the ESO MIDAS
package.
A two dimensional bias correction was done for the object data
as well as for the dome flats. After flatfield correction,
the spectra were extracted from the two dimensional images
with respect to the sky background.
Finally we performed the wavelength calibration using the internal
ThAr lamp of the spectrograph.

\begin{figure}[t]
\begin{center}
\vspace{14cm}
\includegraphics{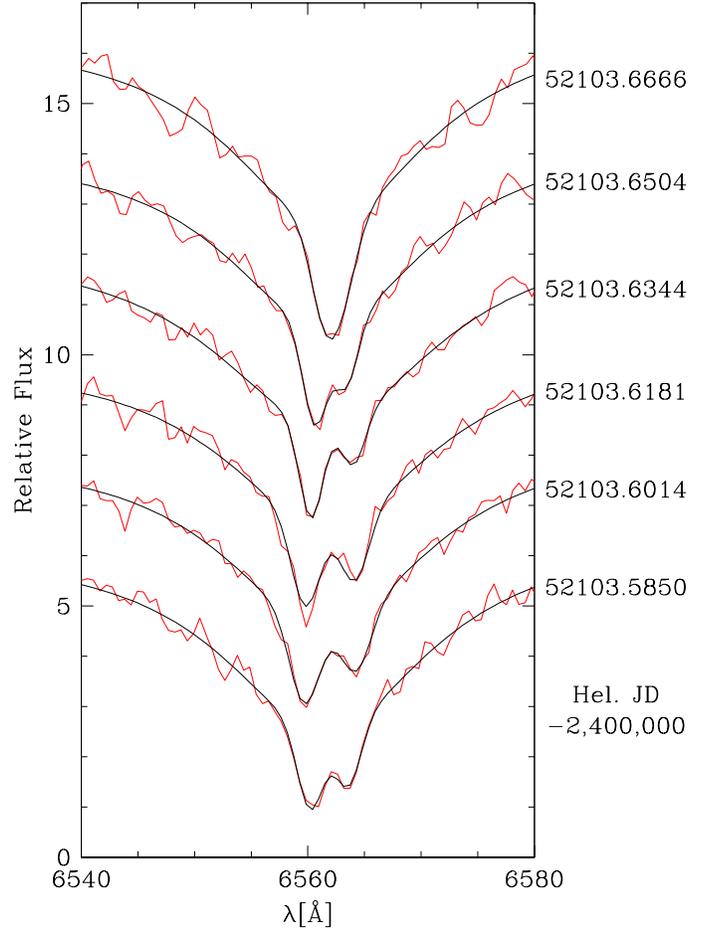}
\caption[]{\label{wandering_lines}
  Spectra of HE\,2209$-$1444 taken at the 3.5\,m telescope of the 
  Calar Alto Observatory during one night. The observed spectra
  are as well plotted as the fits.
  The JD is heliocentric
  corrected and computed for the mid of the exposure.
}
\end{center}
\end{figure}

Figure \ref{wandering_lines} shows a sequence of six
H$_\alpha$ spectra taken at Calar Alto Observatory
during two hours.
The observed spectra are plotted as well as the fits
(cf. Sect. \ref{radvel_section}). The rapid change of the spectral appearance
due to the orbital motion is obvious.

Our TWIN spectra are supplemented by spectra taken with the ESO VLT and
the Isaac Newton Telescope. 
The UVES data we got from the ESO VLT UT2 were taken in December 2000
and October 2001.
Although the data provided by the UVES reduction pipeline were quite useful
for candidate selection, in many cases the quality of the pipeline data
was insufficient for a quantitative analysis of Balmer line profiles.
Therefore, the raw data were reduced with a semi-automatic reduction
package developed in Bamberg (details will be published in a forthcoming
paper).

The last set of data used for our analysis of HE\,2209$-$1444
was taken on October\,28$^{\rm th}$ to October\,30$^{\rm th}$
using the 2.5\,m Isaac Newton Telescope (INT) on La Palma. Eight spectra
were taken with the Intermediate Dispersion Spectrograph (IDS), using the
500\,mm camera and R1200R grating, giving a dispersion of 0.37\,\AA/pix from 6300
to 6700\,\AA {} at a resolution of 1\,\AA.  The spectra were reduced (using the
overscan bias level and a tungsten flat field) with standard
IRAF\footnote{IRAF is distributed by the National Optical Astronomy
  Observatories, which are operated by the Association of Universities for
  Research in Astronomy, Inc., under cooperative agreement with the National
  Science Foundation} tasks, using optimal extraction (Horne\ \cite{Hor86}).
The wavelength calibration was done using CuAr and CuNe lamp taken at the same
position of the telescope and extracted at the same place on the CCD as the
object spectra, typically giving an rms scatter less than 0.01\,\AA.

\section{Radial velocity curve and orbital parameters}\label{radvel_section}

RVs were determined from the TWIN and UVES spectra 
from a fit of a set of mathematical functions to 
the observed line profiles within the MIDAS context. We concentrated
on a 100\,\AA {} wide range of the spectra centered on H$_\alpha$.
Two Gaussians were used to fit both line cores, a Lorentzian to model the line
wings and a linear function to reproduce the overall spectral trend.
Radial velocities of both components were calculated from the
central wavelengths of the fitted Gaussians.
Afterwards the measured RVs were corrected to heliocentric values.  
The INT spectra were analyzed using the MOLLY software package, fitting three
Gaussians to the central part of the H$_\alpha$ line.

\begin{figure}[t]
\epsfxsize95mm
\epsffile{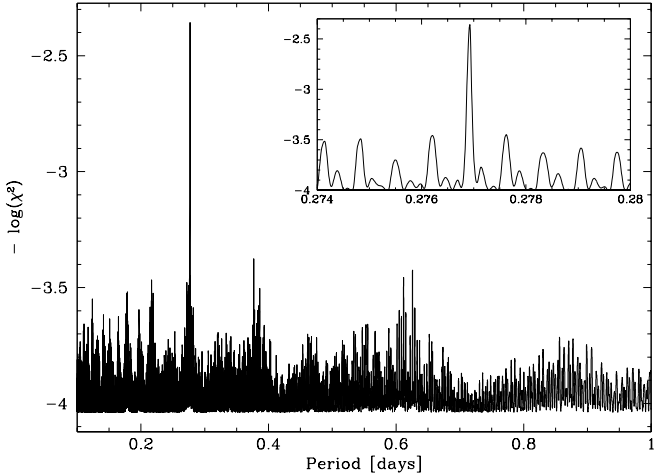}
\caption[]{\label{power}
Combined power spectrum of the system HE\,2209$-$1444. The $\chi^2$ values
obtained from the individual power spectra of the individual 
components are coadded.
}
\end{figure}

\begin{figure}[t]
\epsfxsize95mm
\epsffile{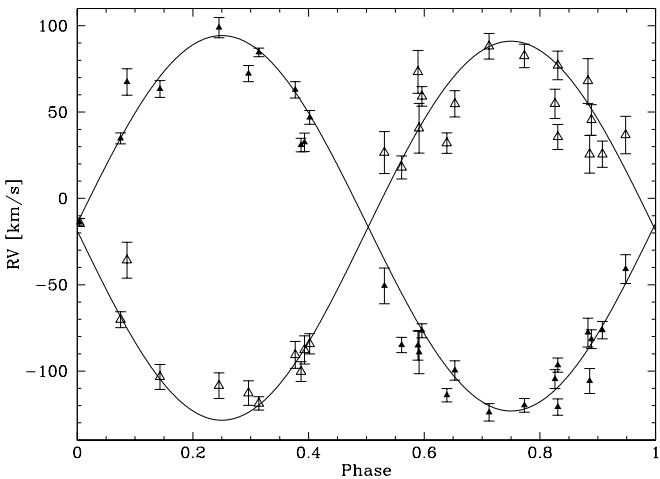}
\caption[]{\label{phase}
Radial velocity curve of the system HE\,2209$-$1444. The measured RVs
for component A (filled triangles) and B (open triangles) are plotted
against the orbital phase.
}
\end{figure}

Both components are very similar, but the left one in Figure~\ref{wandering_lines}
is slightly deeper and broader. This component is called component A further
on. However, sometimes it was difficult to do an unambiguous identification of
the components especially near conjunction phases and in some spectra
with relatively low S/N ratio and
of lower re\-so\-lu\-tion (from Calar Alto and La Palma).
Therefore, we computed the separation of the components first
and fitted sine curves
for a range of periods to the measured RV differences
and determined the $\chi^2$ value for each period. All times
were heliocentric corrected and calculated for the center of the exposure.
Doing this we produced a "difference power spectrum"
indicating the quality of our RV fit
as a function of period (like the one presented in Fig.~\ref{power}).
This way we got
a first estimate of the period of the system and
were subsequently able to identify
the individual components unambiguously for all of our spectra.
This enabled us to compute for each individual component
a "power spectrum".
Both power spectra showed an outstanding peak at
$6^{\rm h}38^{\rm m}46.2^{\rm s}$ (0.276924$^{\rm d}$).
An inspection of the phased RV curves created using periods
corresponding to other peaks in the power spectra allowed us to
rule out aliases, because of one or more strongly deviating
RV measurements.
This way, the period corresponding to the main peak of the power spectrum
remained as an unambiguous solution for component A as well as for
component B -- exactly as it should be for a binary system
(see Fig.~\ref{phase}). 
Finally, adding the $\chi^2$ values of the two
individual power spectra we produced a
"combined power spectrum" for the whole system (Fig.~\ref{power}).

In a final step we determined accurate orbital parameters from a simultaneous
fit of all spectra with the program {\sc fitsb2} (see
Sect.~\ref{s:analysis} and Napiwotzki et al.\ \cite{N03}) in a fashion
similar to the method described in Maxted et al.\ (\cite{M01}). 
Two model profiles,
one for each star, consisting of a combination of two Gaussians for the line 
core and a Lorentzian for the line wings were fitted. 
The position of each profile is
determined from
\begin{displaymath}
\mathrm{RV} _{\mathrm{A/B}} = 
\gamma_{\mathrm{A/B}} \pm K_{\mathrm{A/B}} \sin \left(\frac{T-T_0}{P}\right) .
\end{displaymath}
Free parameters are the mean velocities (including gravitational redshift)
of each component $\gamma_{\mathrm{A/B}}$,
the (projected) orbital velocities $K_{\mathrm{A/B}}$, the zeropoint
$T_0$, and the period $P$, plus the parameters defining the line profiles. 

The final ephemeris of the system for the
time $T_0$ defined as the conjunction time at
which star A moves from the blue side of the RV curve to the red one
(i.e. star A is closest to the observer) is
\begin{eqnarray} 
{\rm Hel.JD}(T_0) = (2,452,097.0349\pm 0.0021) + \nonumber \\ 
    \hspace{37mm}    (0.276928\pm 0.000006) \times E .
\label{ephemeris}
\end{eqnarray}

Also from the RV curves (see Fig.~\ref{phase}) of the {\sc fitsb2} analysis
we found the semi-amplitude of component A to be $K_{\mathrm{A}} = 
108.7 \pm 7.5$\,km/s ($\gamma_{\mathrm{A}} = -14.4 \pm 4.5$ km/s)
whereas for component B the semi-amplitude
is $K_{\mathrm{B}} = 109.1 \pm 10.75$\,km/s 
($\gamma_{\mathrm{B}} = -18.7\pm 6.8$\,km/s).
The mass ratio can be computed from the ratio of the semi-amplitudes to be
\begin{eqnarray}
\frac{M_{\rm B}}{M_{\rm A}} = \frac{K_{\rm A}}{K_{\rm B}} = 1.00 \pm 0.13
\label{massratio}
\end{eqnarray}
i.e the masses of both white dwarfs are identical within the error limits.
For white dwarfs the gravitational redshift can easily be measured from high
resolution spectra (e.g. Reid\ \cite{R96}). As described in paper\,II for
the case of HE\,1414$-$0848, in a double degenerate system the
difference of gravitational redshifts can be measured from $\gamma_{\rm A} -
\gamma_{\rm B}$. If this difference is non zero we obtain an additional
constraint equation, which allows to solve for the individual masses.
However, since the masses of both components
of HE\,2209$-$1444 are very similar, the redshift difference is zero
within error limits. 
Hence we cannot solve for the individual masses from the radial velocity
curve alone. The required additional information, however, will 
be obtained from
a quantitative spectral analysis described in the next section.

\begin{table*}
\begin{center}
\caption[]{Fitted heliocentric radial velocities. The heliocentric corrected JDs
are given for the center of the exposures. The phases $\Phi$ are computed from
the ephemeris given in Eq.~\ref{ephemeris}. At $\Phi$\,=\,0.982
and $\Phi$\,=\,0.040 the individual RVs of both components can not be
measured accurately, because the spectra were taken close to conjunction.
However, these spectra were used to determine the orbital parameters using
the procedure described in Sect.~\ref{radvel_section}.
}
\label{radvel}
\begin{tabular}{c|crrcc} \hline \hline
Hel.JD  & $\Phi$ & \multicolumn{2}{c}{Heliocentric RV
[km/s]} & Telescope & Instrument \\
-2,400,000 & { } & \multicolumn{1}{c}{A} & \multicolumn{1}{c}{B} & { } & { } \\ \hline
51885.5488 & 0.314 & 85 $\pm$ 3 & $-$119 $\pm$ 4 & ESO UT2 & UVES \\
52097.5404 & 0.826 & $-$105 $\pm$ 6 & 55 $\pm$ 9 & CA 3.5\,m &TWIN \\
52097.6125 & 0.086 & 67 $\pm$ 8 & $-$36 $\pm$ 10 & CA 3.5\,m & TWIN \\
52101.5702 & 0.377 & 63 $\pm$ 5 & $-$91 $\pm$ 8 & CA 3.5\,m & TWIN \\
52101.6309 & 0.596 & $-$77 $\pm$ 4 & 59 $\pm$ 6 & CA 3.5\,m & TWIN \\
52102.5410 & 0.883 & $-$78 $\pm$ 8 & 68 $\pm$ 13 & CA 3.5m & TWIN \\
52102.5685 & 0.982 & \multicolumn{1}{c}{------} &
\multicolumn{1}{c}{------} & CA 3.5\,m & TWIN \\
52102.5845 & 0.040 & \multicolumn{1}{c}{------} &
\multicolumn{1}{c}{------} & CA 3.5\,m & TWIN \\
52103.5514 & 0.531 & $-$51 $\pm$ 10 & 27 $\pm$ 12 & CA 3.5\,m & TWIN \\
52103.5678 & 0.591 & $-$89 $\pm$ 12 & 41 $\pm$ 14 & CA 3.5\,m & TWIN \\
52103.5850 & 0.653 & $-$100 $\pm$ 6 & 55 $\pm$ 8 & CA 3.5\,m & TWIN \\
52103.6014 & 0.712 & $-$124 $\pm$ 5 & 88 $\pm$ 7 & CA 3.5\,m & TWIN \\
52103.6182 & 0.773 & $-$120 $\pm$ 4 & 83 $\pm$ 7 & CA 3.5\,m & TWIN \\
52103.6344 & 0.831 & $-$97 $\pm$ 4 & 77 $\pm$ 8 & CA 3.5\,m & TWIN \\
52103.6504 & 0.889 & $-$82 $\pm$ 5 & 45 $\pm$ 9 & CA 3.5\,m & TWIN \\
52103.6666 & 0.948 & $-$41 $\pm$ 8 & 37 $\pm$ 11 & CA 3.5\,m & TWIN \\
52104.6750 & 0.589 & $-$85 $\pm$ 8 & 73 $\pm$ 12 & CA 3.5\,m & TWIN \\
52105.6591 & 0.143 & 63 $\pm$ 5 & $-$103 $\pm$ 7 & CA 3.5\,m & TWIN \\
52106.5183 & 0.245 & 99 $\pm$ 6 & $-$108 $\pm$ 7 & CA 3.5\,m & TWIN \\
52172.6372 & 0.004 & $-$14 $\pm$ 2 & $-$14 $\pm$ 2 & ESO UT2 & UVES \\
52194.5343 & 0.075 & 35 $\pm$ 3 & $-$70 $\pm$ 5 & ESO UT2 & UVES \\
52194.6249 & 0.402 & 47 $\pm$ 4 & $-$84 $\pm$ 6 & ESO UT2 & UVES \\
52211.3592 & 0.831 & $-$121 $\pm$ 5 & 36 $\pm$ 7 & INT & IDS \\
52211.3807 & 0.908 & $-$76 $\pm$ 5 & 26 $\pm$ 8 & INT & IDS \\
52212.3189 & 0.296 & 72 $\pm$ 5 & $-$112 $\pm$ 7 & INT & IDS \\
52212.3440 & 0.387 & 31 $\pm$ 4 & $-$100 $\pm$ 6 & INT & IDS \\
52212.3921 & 0.561 & $-$85 $\pm$ 4 & 18 $\pm$ 7 & INT & IDS \\
52212.4138 & 0.639 & $-$114 $\pm$ 4 & 32 $\pm$ 6 & INT & IDS \\
52212.4822 & 0.886 & $-$106 $\pm$ 7 & 26 $\pm$ 11 & INT & IDS \\
52213.4533 & 0.393 & 32 $\pm$ 5 & $-$88 $\pm$ 8 & INT & IDS \\ \hline
\end{tabular}
\end{center}
\end{table*}

\section{Spectroscopic analysis} \label{s:analysis}

Since in the case of HE\,2209$-$1444 mass estimates from the RV curves
alone are impossible, we had to rely on a model atmosphere analysis.
Because this system is double-lined the
spectra are a superposition of both individual white dwarf spectra.
This problem is sometimes circumvented by assuming that the results of
a simple single-lined model fit is representative for the average
parameters.
In paper~II we estimated the mean gravity and individual
temperatures of the components of HE\,1414$-$0848 from the analysis of
individual spectra taken close to conjunction and quadrature phases.
However, this approach is of limited accuracy, only. 

A direct approach would be to disentangle the observed spectra
by deconvolution techniques into the spectra of the individual components.
Then we could analyze the spectra by fitting
synthetic spectra developed for single-lined white dwarfs
to the individual line profiles.
Such a disentangeling procedure has
been developed by Simon and Sturm\ (\cite{S94}) and was sucessfully applied
to main sequence double-lined binaries.
However, it has not been tested for
white dwarfs, for which the wavelength shifts caused by orbital motions are
much smaller than the line widths of the broad Balmer lines.
Therefore we choose a different approach for our analysis of
the HE\,2209$-$1444 system. We used
the program {\sc fitsb2} (Napiwotzki et al.\ \cite{N03}), which
performs a spectral analysis of both components of double-lined systems.
It is based on a $\chi^2$ minimization technique using a simplex algorithm.
The fit is performed on all
available spectra covering different spectral phases simultaneously.
For HE\,1414$-$0848 (paper~II) we made use of spectra taken at conjunction
or quadrature only. But now, all available
spectral information is combined into the parameter determination procedure.
An application of the program {\sc fitsb2} to
HE\,1414$-$0848 yielded encouraging results (Napiwotzki et al.\ \cite{N03}).
A detailed description of the program and an evaluation of
its performance will follow in a subsequent
paper (Napiwotzki, in prep.). 

A large grid of synthetic spectra for DA white dwarfs 
computed from LTE model atmospheres with a code
described in Finley et al.\ (\cite{F97}) was used for the analysis. 
A simultaneous fit of the Balmer lines H$_{\beta}$ to H$_8$ (UVES spectra)
or H$_{\beta}$ to H$_{\delta}$ (TWIN spectra) was performed.
We did not include the INT spectra for the model atmosphere analysis, because
these spectra cover only the H$_\alpha$ range.
For details refer to
Koester et al.\ (\cite{K01} and references therein). 
The model spectra were convolved with Gaussians with FWHMs corresponding to
the resolution of the observed spectra. 

The total number of
fit parameters (stellar and orbital) is high. Therefore we fixed as many
parameters as possible before performing the model atmosphere analysis.
We have kept the radial velocities of the
individual components fixed according to the radial
velocity curve presented in Sect.~\ref{radvel_section}.
Since the mass ratio is already accurately determined from the radial
velocity curve we fixed the gravity ratio.
The remaining fit parameters are the effective
temperatures of both components and the gravity 
of the primary.
The gravity of the secondary is adjusted according to the
primary value during the fitting procedure.
The surface gravities also determine the relative
weight of the two model
spectra from the radius, obtained from the mass-radius relation of
Benvenuto \& Althaus\ (\cite{B99}).
The flux ratio in the V-band is calculated from the actual
parameters and the model fluxes are scaled accordingly.
The individual contributions are updated
consistently as part of the iteration procedure.

\begin{figure*}[t]
\epsfxsize19cm
\epsffile{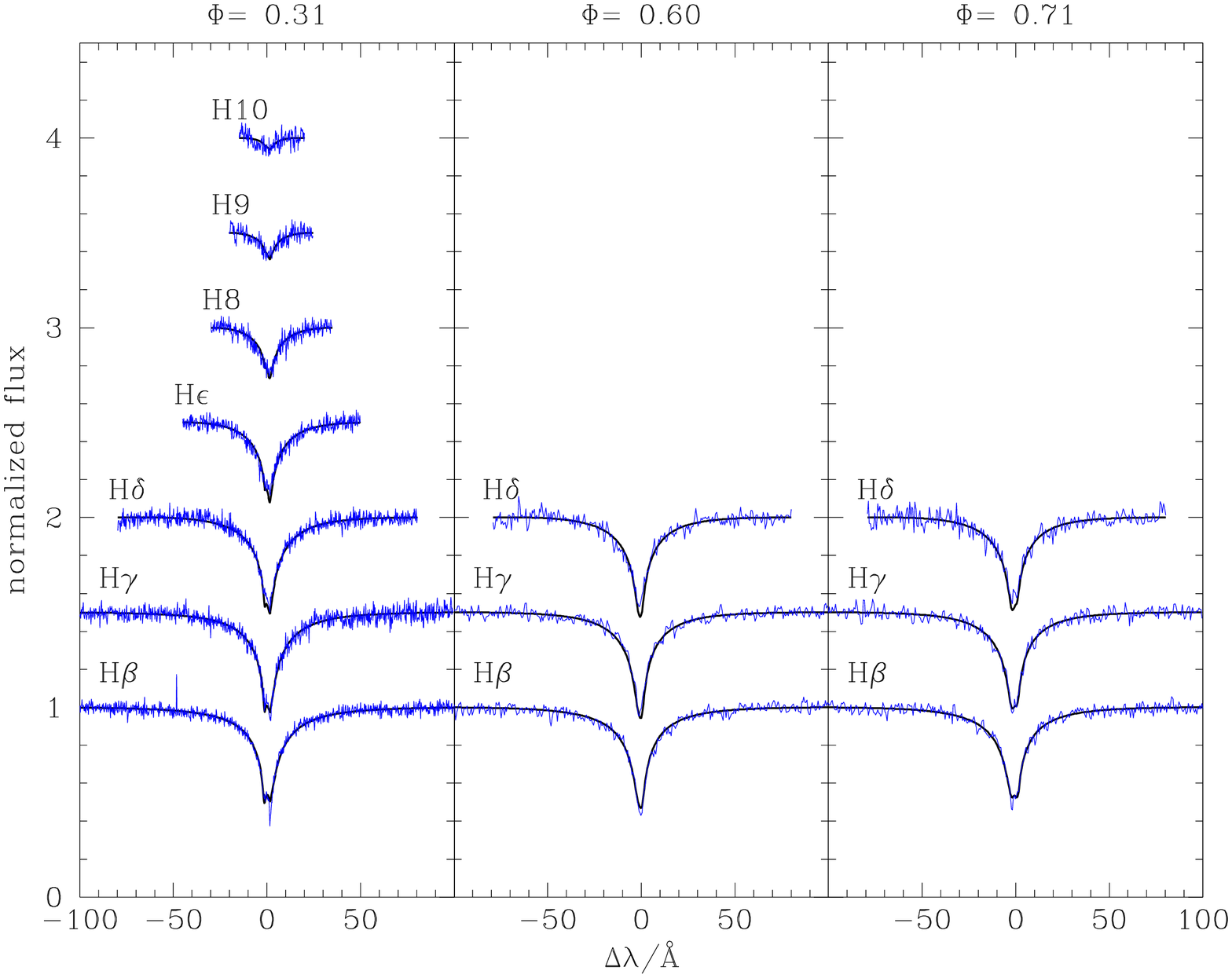}
\caption[]{\label{f:he2209fit}
  Sample Model atmosphere fits of HE\,2209$-$1444 phase spectra taken at
  different phases: The discovery spectrum ($\Phi$\,=\,0.31) taken
  with the UVES spectrograph and two TWIN spectra ($\Phi$\,=\,0.60 and
  $\Phi$\,=\,0.71).
}
\end{figure*}

Strong NLTE cores are present in H$_{\alpha}$ and H$_{\beta}$, which cannot be
reproduced by our LTE model spectra.
In principle, these lines are important for the
temperature determination, especially of the cooler and fainter B
component. However,
due to the large NLTE effects we had to exclude H$_\alpha$ completely and
the core of H$_\beta$ ($\pm$ 4\,\AA). The final
fit results are summarized in Table~\ref{t:fitsb2fit} and a sample fit is
shown in Fig.~\ref{f:he2209fit}. The fit of the line cores is not perfect,
even after excluding H$_\alpha$. The parameter which is most sensitive to
details of the fitting procedure is, as may be expected, the
temperature of the fainter component B. If, e.g., we include the wings of the
H$_\alpha$ line we increase $\teff$ of component B by almost 500\,K, while
the temperature of A and the gravity are only modified within the error
limits. Thus, while the temperature of B is less well determined than that of
A, the basic
properties of the HE\,2209$-$1444 system are reliable.
Both components have identical masses
(M$_{\rm A}$\,=\,M$_{\rm B}$\,=\,0.58\,M$_{\odot}$)
but different temperatures (8490\,K vs.\ 7140\,K).
White dwarf masses were computed from the mass\,-\,radius
relations of Benvenuto \& Althaus\ (\cite{B99})
with ``thick hydrogen
envelopes'' (M$_{\rm H}$/M$_{\rm WD}$ = 10$^{-4}$). This result does not
depend much on the choice of a particular model computation. 
From the thick envelope
cooling sequences of Wood (\cite{W95}) we derived virtually identical
results: 0.58\,M$_{\odot}$ and 0.57\,M$_{\odot}$, respectively. However,
since the HE\,2209$-$1444 system is the result of a common envelope
evolution it is not clear which envelope layer is the correct one. Using
the ``thin'' envelope models (M$_{\rm H}$\,=\,0) of Benvenuto \& Althaus\
(\cite{B99}) as the other extreme would yield slightly lower masses
(0.57\,M$_{\odot}$ and 0.56\,M$_{\odot}$, resp.).
The resulting sum of masses of the 
HE\,2209$-$1444 system is $1.15\pm 0.07 M_\odot$. The error limit for the
sum of masses 
is smaller than expected from a simple combination of the individual errors,
because the mass errors of the individual components are partially 
anti-correlated.

\begin{table}
\caption{Results of the model atmosphere analysis with {\sc fitsb2}. The
error limits include the uncertainties of orbital parameter determination.
\label{t:fitsb2fit}}
\begin{center}
\begin{tabular}{l|ccc|r}
comp.&$\teff$/K&$\log g$&$M/M_\odot$&age/Gyrs\\ \hline
A & $8490\pm 80$  & $7.97\pm 0.05$ & $0.58\pm 0.03$ & 0.9\\
B & $7140\pm 110$ & $7.97\pm 0.13$ & $0.58\pm 0.08$ & 1.4
\end{tabular}
\end{center}
\end{table}

\section{Evolutionary status} \label{discussion}

From the temperatures derived above we can estimate the ages of both
components using cooling tracks for white dwarfs. Our estimate is
based on the calculations of Benvenuto \& Althaus\ (\cite{B99}) for
C/O white dwarfs with thick hydrogen envelopes.  Ages were
interpolated from these tracks for the derived masses and temperatures 
of the white dwarfs.
Resulting ages are 0.9\,Gyrs for A and 1.4\,Gyrs
for B, i.e.\ the cooling ages of both white dwarfs differ by 0.5\,Gyrs
or 50\%.

\begin{figure}[t]
\epsfxsize95mm
\epsffile{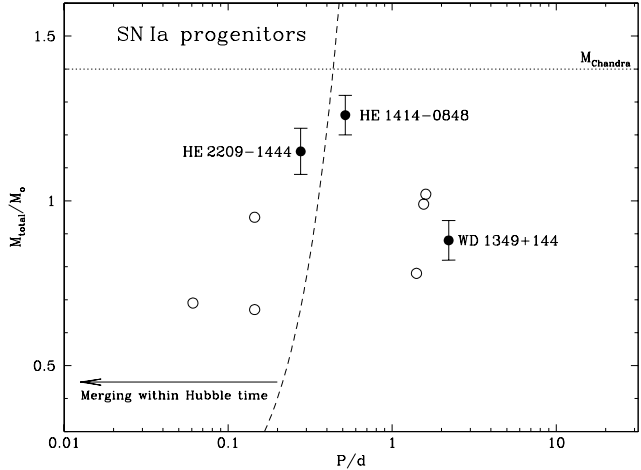}
\caption[]{\label{f:pmsample}
Mass\,-\,period diagram of known double-lined DD systems with mass
determinations for both components. Literature values from Maxted et al.
(\cite{M99} and \cite{MMM00c}) are plotted as open
circles while SPY results are shown as filled circles with error bars
(WD\,1349\,$+$\,144 parameters from Karl et al. \cite{K03}). The
Chandrasekhar limit and the region of DDs, which will merge within a Hubble
time are indicated.
}
\end{figure}

The separation between both white dwarfs is quite small, only $1.87
R_\odot$. Thus HE\,2209$-$1444 obviously underwent phases of strong
binary interaction in its history. From the size of the orbits and the
period the orbital velocity can be computed to be 171\,km/s.
The comparison
with the observed RV amplitudes allows us to determine the inclination
of this system as $i=40^\circ$. This system will merge within 5\,Gyrs
due to gravitational wave radiation.
Figure~\ref{f:pmsample} shows the total masses and periods of
the sample of known double-lined DD systems.
As can be seen, neither HE\,2209$-$1444 nor HE\,1414$-$0848 (Paper II)
qualify as a SN Ia progenitor.
However, both objects found by SPY are much closer
to the region where such progenitors are expected than any
DD system known before.

\begin{figure} 
\epsfxsize95mm
\epsffile[40 32 701 480]{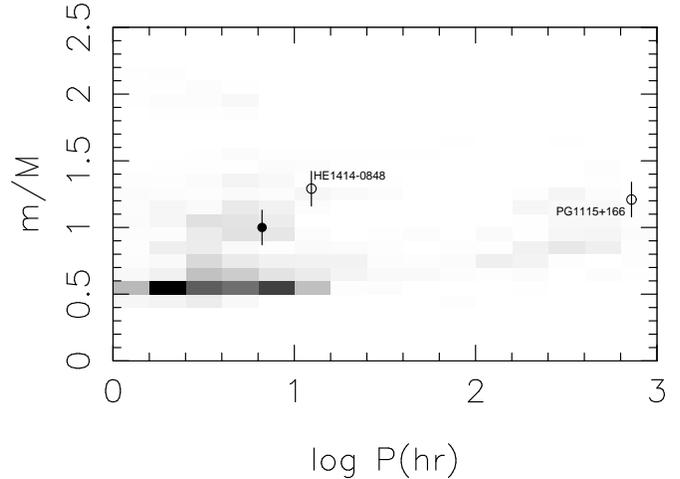}
\caption[]{Period\,-\,mass ratio distribution for the current population of DDs
  in the Galaxy for systems in which at least one white dwarf has a mass
  above 0.5\,M$_\odot$ from model A2 of Nelemans et al.\ (\cite{NYP01}).
  Only systems with a luminosity ratio between the components of less than 5 are
  plotted.
  Grey shades indicate the expected number of systems brighter than
  V\,=\,$15.5^{\mathrm{m}}$.
  The derived period and mass ratio of HE\,2209$-$1444
  is plotted as a filled dot with an error bar.
  Also plotted are HE\,1414\,-\,0848 (paper II) and PG\,1115\,+\,166
  (Bergeron et al., \cite{B2002} and Maxted et al., \cite{M2002}).}
\label{f:pq}
\end{figure}


We can compare the properties of HE\,2209$-$1444 with the outcome of
computations for the synthesis of the population of double white
dwarfs of Nelemans et al.\ (\cite{NYP01}).
Fig.~\ref{f:pq} shows the expected period\,-\,mass ratio distribution of systems
in which one component has a mass 
above 0.50\,M$_\odot$  
from model A2 of 
Nelemans et al.\ (\cite{NYP01}).
Only systems with a luminosity ratio between the 
components of 
less than 5
are plotted. The derived period and mass ratio of 
HE\,2209$-$1444 is marked in Fig.~\ref{f:pq} with its error bar.
The figure shows that such systems are expected to have
periods smaller than $\approx$10 hours and either similar or
more massive companions.

HE\,2209$-$1444 confirms the trend that double white dwarfs tend to
have roughly equal masses, in contrast to population models using a
standard common envelope description (see for a discussion Maxted et
al.\ \cite{MMM00c}, Nelemans et al.\ \cite{NYP01}) and more in line
with the results presented in Nelemans et al.\ \cite{NYP01},
in which
the outcome of dynamical unstable mass transfer is calculated from the
angular momentum balance, rather than the energy balance. However, it
should be noted that a model using standard common envelope also
produces systems similar to HE\,2209$-$1444.

The SPY project will provide a large, homogeneous sample of close
double white dwarfs, which will enable us to confirm or refuse the
alternative common envelope formalism.

\section{Summary} \label{summary}

We report the discovery of the close double-lined binary
HE\,2209$-$1444 by SPY. From follow-up observations at three different
telescopes we derived the radial velocity curve. The analysis yielded
an orbital period of $P = 6^{\rm h}38^{\rm m}47^{\rm s}$
and identical semi-amplitudes of 109\,km/s for both components. Parameters of
both individual components were determined from a model atmosphere
analysis. We made use of a novel method which combines all available
spectra covering the different orbital phases for a simultaneous
fit. The result is a common gravity of 7.97 and temperatures of 8490\,K
and 7140\,K for components A and B, respectively.

Masses were determined from a subsequent comparison with the
cooling tracks of Benvenuto \& Althaus\ (\cite{B99}). The results are
masses of $0.58M_\odot$ for the individual components and a total mass
for the HE\,2209$-$1444 system of $1.15\pm 0.07M_\odot$. This makes
HE\,2209$-$1444 the second most massive DD ever found in a RV survey,
surpassed only by HE\,1414$-$0848, another detection of the SPY
project (Napiwotzki et al.\ \cite{N02}).
However, the seperation of the components in HE\,2209$-$1444 is smaller than
in the case of HE\,1414$-$0848. Therefore, the merging time of the
former system is 5\,Gyrs whereas in the latter case the system
will merge within two Hubble times.

We compared the resulting orbital parameters of HE\,2209$-$1444 with
the outcome of theoretical calculations for the formation of DDs and
find that the properties of HE\,2209$-$1444 are in
agreement with recent predictions. 

\acknowledgement C.A.K. is supported by DFG under grant Na365/3-1
and by travel grants. G.N. would like to thank Tom Marsh for
the use of his MOLLY and aliases software.

\end{document}